\documentclass[12pt,a4paper]{article}
\usepackage{amsmath}
\usepackage{amsxtra}
    \usepackage{amstext}
    \usepackage{amssymb}
    \usepackage{latexsym}
    \usepackage{graphicx}
\usepackage{color}
\usepackage{graphics}

\newcommand{\hepth}[1]{{\tt hep-th/#1}}

\newcommand{\B}{\bullet}

\newcommand{\p}{\vspace{6pt}\noindent}
\newcommand{\jump}{\vspace{2pt}}

\advance\textheight by \topskip
\makeatletter

\@addtoreset{equation}{section}
\def\section{\@startsection {section}{1}{\z@}{-8.5ex plus -1ex minus
 -.2ex}{3.3ex plus .2ex}{\large\bf}}%\centering}}
\def\subsection{\@startsection{subsection}{2}{\z@}{-3.25ex plus
 -1ex minus -.2ex}{1.5ex plus .2ex}{\bf}}
\def\subsubsection{\@startsection{subsubsection}{3}{\z@}{-3.25ex plus%
 -1ex minus -.2ex}{1.5ex plus .2ex}{\sl}}

\begin{document}

\begin{titlepage}
\vspace*{-2cm}
\begin{flushright}
%%YITP-05-29
\end{flushright}

\vspace{0.3cm}

\begin{center}
{\Large {\bf }} \vspace{1cm} {\Large {\bf Aspects of defects in
integrable quantum field theory}}\\
\vspace{1cm} {\large  E.\ Corrigan\footnote{\noindent E-mail:
{\tt edward.corrigan@durham.ac.uk}} } \\

\vspace{0.3cm}
{\em Department of Mathematical Sciences \\ University of Durham,
Durham DH1 3LE, U.K.} \\

\vspace{2cm}

\vspace{.5cm}
Invited talk at CQIS, Protvino, Russia\\ January 2011\\
\vspace{1cm}
{\bf Abstract}
\vspace{.5cm}
\end{center}
Defects are ubiquitous in nature, for example dislocations,
shocks, bores,
or impurities of various kinds, and their descriptions
are an important part of any
physical theory. However, one might ask the question: what types
of defect are allowed and what are their properties if it is
required to maintain integrability within an integrable field
theory in two-dimensional space-time? This talk addresses a
collection of ideas and questions including examples of
integrable defects and the curiously special
roles played by energy-momentum conservation and B\"acklund
transformations, solitons scattering with defects and some
interesting
effects within the sine-Gordon model, defects in integrable
quantum field theory and the construction of transmission
matrices,
and concluding with remarks on algebraic considerations and
future directions.

\p

\end{titlepage}

\parindent=0pt

\textbf{\large 1\quad An integrable discontinuity}

\p Start with a single selected point on the {$x$}-axis, say
{$x_0=0$}, and denote the field to the
left of it ({$x<x_0$}) by $u$, and to the right ({$x>x_0$})
by {$v$}:

\vspace{20pt}

\setlength{\unitlength}{1cm}
\ \ \ \ \ \ \ \ \ \ \ \ \ \ \ \ \ \ \ \ \ \ \ \ \
\begin{picture}(0,0)
\put(0,0){$\dots$} \put(.6,0){\line(1,0){4}}
\put(4.55,-.1){$\B$}\put(4.7,0){\line(1,0){4}}
\put(8.85,0){$\dots$}
\put(2,-.5){{$u(x,t)$}}\put(4.55,-.5){{$x_0$}}
\put(6.1,-.5){{$v(x,t)$}}
\end{picture}
\vspace{30pt}

The field equations in the two separated domains are:
\begin{equation}\nonumber
\partial^2 u=-\frac{\partial U}{\partial u},\quad x<x_0,\quad
\partial^2 v=-\frac{\partial V}{\partial v},\quad x>x_0.
\end{equation}
If $U,V$ both represent potentials for integrable models, the
question really concerns how the fields $u,v$ may be `sewn'
together at $x_0$ to preserve integrability? There may be
several kinds of sewing condition and each would represent
a specific type of defect. An elaboration would also allow
some degrees of freedom attached to the point $x=0$ and allowed
to interact with both fields. That possibility will be
considered later. If there is no additional degree of freedom,
the defect will be called `type I'. The fields $u,v$ could
be multicomponent but for most of the time in this talk they
will refer to single component scalar fields. If there is
just one defect it will be placed at $x_0=0$ from now on.

\p One natural choice (a {$\delta$}-impurity) would be
to have the two fields continuous across $x=0$ but allow
a discontinuity in the first spatial derivative. For example,
put
$${u(0,t)=v(0,t),\quad
u_x(0,t)-v_x(0,t)=\mu \, F(u(0,t)))},$$
but, while there are likely to be interesting effects the
integrability is actually lost \cite{ghw2002}.
%\end{frame}
%
To try to make some progress, note there is clearly a
distinguished point, translation symmetry is lost and the
conservation laws - at least some of them - (for example,
energy-momentum), are violated unless the
impurity contributes compensating terms. The problem can
be restated in terms of seeking the suitable compensating
associated with the defect to ensure there are enough
conserved quantities to guarantee integrability.

\p Consider first the field contributions to energy-momentum:
$${{P^\mu}=\int_{-\infty}^0 dx\, T^{0\mu}(u) +
\int^{\infty}_0 dx\, T^{0\mu}(v)}.$$
Using the field equations, can we arrange
$${\frac{ dP^\mu}{dt}= -\left[T^{1\mu}(u)\right]_{x=0} +
\left[T^{1\mu}(v)\right]_{x=0}}
  ={-\frac{d{ D}^\mu(u,v)}{dt}},$$
with the right hand side depending only on the fields at $x=0$?
If so, ${{ P}^\mu +{ D}^\mu}$ will be conserved with
${{ D}^\mu}$ being the defect contribution.
Only a few possible sewing conditions (and bulk potentials
${U,\ V}$)  are permitted for this to work.

\p Note, it might be more natural to consider higher spin
charges and indeed that was the historical route
\cite{bczlandau}. However, for reasons that are not yet
entirely clear it appears to be enough to investigate the
conservation of momentum alone, since that already produces
constraints equivalent to those obtained by other means.

\p Consider first the field contributions to energy and
calculate
$${\frac{ dP^0}{dt}=[u_xu_t]_0 - [v_xv_t]_0}.$$
Setting {$u_x=v_t+X(u,v), \ \ v_x=u_t+Y(u,v)$}, clearly
$${\frac{ dP^0}{dt}=u_tX-v_tY}.$$
This is a total time derivative if $${X=
-\frac{\partial D^0}{\partial u}, \ \
Y=\frac{\partial D^0}{\partial v}},$$
for some {$D^0$}.  Then
$${\frac{ dP^0}{dt}=-\frac{dD^0}{dt}}.$$
This was only to be expected, with no further constraints on
$U(u), V(v)$ or $D^0$, since time translation has not been
deliberately violated.

\p On the other hand, for momentum
$${\frac{ dP^1}{dt}= -\left[\frac{u_t^2+u_x^2}{2}-U(u)\right]_{x=0}
+\left[\frac{v_t^2+v_x^2}{2}-
V(v)\right]_{x=0}}$$
$${=\left[-v_tX +u_tY -\frac{X^2-Y^2}{2}+U-V\right]_{x=0}}=
{-\frac{d D^1}{dt}.}$$
This is a total time derivative of a functional of the fields
at $x=0$ provided the first piece is a perfect differential
and the second piece vanishes. Thus
$${X=-\frac{\partial D^0}{\partial u}=
\frac{\partial D^1}{\partial v},\ \
Y=\frac{\partial D^0}{\partial v}=
-\frac{\partial D^1}{\partial u}},$$
and
$${\frac{\partial^2 D^0}{\partial v^2}=
\frac{\partial^2 D^0}{\partial u^2},\quad
\frac{1}{2}\left(\frac{\partial D^0}{\partial u}
\right)^2-\frac{1}{2}
\left(\frac{\partial D^0}{\partial v}
\right)^2=(U-V)},$$
which is strongly constraining. What are the possible
combinations {$U,V,D^0\,$}? It is not difficult to check
that sine-Gordon, Liouville, massless free and massive
free are in fact the only possibilities.
For example, if {$U(u)=m^2u^2/2,$ $V(v)=m^2v^2/2$},
{$D^0$} turns out to be
$${D^0(u,v)=\frac{m\sigma}{4}(u+v)^2 +
\frac{m}{4\sigma}(u-v)^2},$$
and {$\sigma$} is a free parameter.

\p Note though, the other single field integrable
theory of a similar type, the Tzitz\'eica potential
(otherwise known as Bullough-Dodd, Mikhailov-Zhiber-Shabat,
or {$a_2^{(2)}$} affine Toda),
$${U(u)=e^u+2e^{-u/2}},$$ is not a possible solution.
This was mysterious for a while and the resolution will
come later when a more general type of defect will be explored.

\p Note also,there is a Lagrangian description of this
type of defect (type I) \cite{bczlandau}:
{\begin{eqnarray}\nonumber
{\cal L}&=&\theta(-x){\cal L}(u)+\delta(x)\left(
\frac{uv_t-u_t v}{2} - D^0(u,v)\right)
+\theta(x){\cal L}(v).
\end{eqnarray}}

\vspace{5pt}
\textbf{\large 2\quad sine-Gordon}

\p Choosing {$u,v$} to be sine-Gordon fields (and scaling
the coupling and mass parameters to unity),
then:
$${D^0(u,v)=2\left(\sigma\cos\frac{u+v}{2} +
\sigma^{-1}\cos\frac{u-v}{2}\right)}$$
to find
\begin{eqnarray*}
&&x<x_0:\quad  \partial^2 u = -\sin u;\qquad x>x_0:
\quad \partial^2 v = -\sin v,  \\
&&x=x_0:\ \ \quad   u_x= v_t-\sigma\sin\frac{u+v}{2}-
\sigma^{-1}\sin\frac{u-v}{2},\\
&&x=x_0:\ \ \quad  v_x= u_t+\sigma\sin\frac{u+v}{2}-
\sigma^{-1}\sin\frac{u-v}{2}.
\end{eqnarray*}
Unexpectedly, the last two expressions are recognised
to be a B\"acklund transformation `frozen' at {$x=0$},
already a signal of integrability. In fact, in this
case a Lax pair can be devised to incorporate the defect
and generate suitably modified higher spin conserved
charges. So the system is integrable in the standard
way \cite{bczlandau}.

\p Consider a soliton incident from {$x<0$}. It will
not be possible
to satisfy the sewing conditions (in general) unless
a similar soliton
emerges into the region { $x>0$}. Thus, using convenient
expressions for the standard soliton solutions
{\begin{eqnarray*}\nonumber
&&e^{iu/2}=\frac{1+iE}{1-iE},\ \ e^{iv/2}=
\frac{1+izE}{1-izE},\ \ E=e^{ax+bt+c},
\ \  a=\cosh\theta,\ \ b=-\sinh\theta,
\end{eqnarray*}}
where { $z$} is to be determined. It is also useful to
set {$\lambda=e^{-\eta}$} and then find the sewing
conditions imply
 $$ {z=\coth\left(\frac{\eta-\theta}{2}\right)}.$$
This result has several interesting consequences
\cite{bczlandau, bcz2005}:
\begin{itemize} \item {$\theta >\eta$} implies {$z<0$},
meaning the soliton emerges as an anti-soliton
 and the final state will contain a discontinuity of
 magnitude {$4\pi$} at {$x=0$};
\item {$\theta=\eta$} implies {$z=\infty$} and there
is no emerging soliton, meaning
the energy-momentum of the soliton is captured by
the `defect', as is its topological charge indicated
by a discontinuity {$2\pi$} in the final state;
\item  {$\theta<\eta$} implies  {$z>0$} meaning the
soliton retains its character, and the final state
does not acquire an additional discontinuity.
\end{itemize}
It's worth making a few other comments:

\begin{itemize} \item Defects at {$x=x_1< x_2< x_3<
\dots <x_n$} behave independently, each contributes a
factor
{$z_i$} for a total `delay' {$z=z_1 z_2 \dots z_n$};
\item Each component of a multisoliton is affected separately,
thus at most one of them can be `filtered out' (because the
rapidities of individual soliton components must be different);
\item Since a soliton can be absorbed, can a starting configuration
with {$u=0$, $v=2\pi$}
decay into a soliton? This would seem to need quantum mechanics
to provide a probability for decay after a specific time.
\item The properties listed above suggest that a type I defect
could be used to manipulate solitons, see for example  \cite{cz2004}.
\item Extending the type I defect to other field theories of
affine Toda type is only possible for the series of models based
on $a_n^{(1)}$ \cite{bcz2004, cz2007} .
\end{itemize}

\vspace{10pt}
\textbf{\large 3 \quad Classical type II defect}

\p Since not all field theories of affine Toda type can
accommodate a type I defect it is natural to try adding
extra degrees of freedom to the defect itself \cite{cz2009}.
For example, consider two relativistic field theories with
fields {$u$} and {$v$}, and add {$\lambda(t)$} at the defect
location as follows:
$${{\cal L}=\theta(-x){\cal L}_u + \theta(x){\cal L}_v +
\delta(x)\left(2 q\lambda_t -D^0(\lambda,p,q)\right),}$$
where
$${q=\left. \frac{u-v}{2}\right|_0,\qquad p=\left.
\frac{u+v}{2}\right|_0}.$$

Then the usual steps lead to equations of motion:
$${\partial^2 u=-\frac{\partial U}{\partial u},
\quad x<0,\qquad \partial^2 v=-\frac{\partial V}{\partial v},
\quad x>0},$$
alongside defect conditions at {$x=0$}:
$${2 q_x=-D^0_p,\qquad 2p_x-2\lambda_t=-D^0_q,\qquad
2q_t=-D^0_\lambda}.$$
Note, if there was no defect potential $\lambda$ would
act as a lagrange multiplier.

\p As before, consider momentum
$${P^1=-\int_{-\infty}^0 dx\, u_t u_x-\int^{\infty}_0
dx\, v_t v_x},$$
and seek a functional {$D^1(u,v,\lambda)$} such that
{$ P^1_t\equiv-D^1_t$}. Then {$P^1+D^1|_{x=0}$}
will be the total conserved momentum of the system.
This leads to the following constraints on {$U,\, V,\, D^1$}:
$${D^0_p=D^1_\lambda,\qquad D^0_\lambda=D^1_p,\qquad
D^1_\lambda D^0_q-D^1_qD^0_\lambda=2(U-V)},$$
implying
$${D^0=f(p+\lambda,q)+g(p-\lambda,q),\qquad D^1=
f(p+\lambda,q)-g(p-\lambda,q),}$$
with
$$
 {f_\lambda g_q-g_\lambda f_q=U(u)-V(v)}.
$$
The latter is also a powerful constraint since the left
hand side depends on $\lambda$ while the right hand side
does not. However, this time it is not totally clear what
the full set of solutions is. Nevertheless, it is certainly
possible to choose {$f,g$} in such a way that the potentials
{$U,V$} can be any one of
sine-Gordon, Liouville, Tzitz\'eica, or free massive or
massless; ie any one of the full set of single field,
relativistic, integrable field theories. For example,
for Tzitz\'eica:
$$U(u)=(e^{u}+2\,e^{-u/2}-3),\quad
V(v)=(e^{v}+2\,e^{-v/2}-3)$$
the corresponding defect potential is given by
\begin{eqnarray*}\label{defectpotentiala22}
&D^0=&2\sigma\left(  e^{(p+\lambda)/2}+e^{-(p+\lambda)/4}
\,\left(e^{q/2}+e^{-q/2}\right)\right)%\nonumber \\
+\frac{1}{\sigma} \left(8\,e^{-(p-\lambda)/4}+e^{(p-\lambda)/2}
\,\left(e^{q/2}+e^{-q/2}\right)^2\right).
\end{eqnarray*}
But note, the defect conditions following from this are not
a `frozen' B\"acklund transformation. Moreover, other
multi-component affine Toda field theories can support
type II defects, though not all of them (the missing ones
are those based on root data corresponding to Kac-Dynkin
diagrams containing a simple root with three connected
neighbours - such as the $e$-series), which suggests there
might be a type III Lagrangian waiting to be found. In the
sine-Gordon model the type-II defect is also new - in a
sense it is really two `fused' type-I defects - and the
corresponding defect conditions are not quite a frozen
B\"acklund transformation even in that case.

\vspace{15pt}
\textbf{\large 4\quad Defects in quantum field theory}

\p Based on facts gleaned from the classical soliton-defect
scattering the following might be expected:
\begin{itemize}
\item soliton-defect scattering - pure transmission compatible
with the bulk S-matrix and topological charge will be preserved
but may be exchanged with the defect;
\item  for each type of defect there may be several types of
transmission matrix (eg in sine-Gordon expect two different
transmission matrices since the
topological charge on a defect
can only change by {$\pm 2$} when a soliton/anti-soliton passes).
\item  not all transmission matrices need be unitary - in the
sine-Gordon model one turns out to be a resonance of the other.
\end{itemize}
Then, seek to understand different types of defect, assemblies
of defects, defect-defect scattering, fusing defects, and so on.

\p The most relevant quantity will be the `transmission matrix',
denoted,
$$T_{a\alpha}^{b\beta}(\theta, \eta), \quad a+\alpha=b+\beta,$$
where $\alpha,\beta$ and $a,b$ are defect and soliton labels,
respectively, and $\eta$ is a collection of defect parameters.
Besides any general considerations, such as unitarity (where
appropriate), or crossing properties, the transmission matrix
will have to be compatible with the bulk S-matrix, suggesting
the following algebraic constraints \cite{dms1994}:
$${S_{ab}^{cd}(\Theta)\,T_{d\alpha}^{f\beta}(\theta_a)
T_{c\beta}^{e\gamma}(\theta_b)=
T_{b\alpha}^{d\beta}(\theta_b)
T_{a\beta}^{c\gamma}(\theta_a)S_{cd}^{ef}(\Theta)}.$$
Here {$\Theta=\theta_a-\theta_b$} and sums are to be understood
over the `internal', repeated
labels {$\beta,c,d$}.

\p Generally, for affine Toda field theories based on the root
data of a Lie algebra $g$, a number of different transmission
matrices are expected (all of them infinite-dimensional), labelled
by elements of the associated weight lattice differing by roots.
Thus the two possibilities for sine-Gordon, labelled by even or
odd integers, respectively, might be more appropriately considered
as labeled by integer or half-odd-integer spins.

\p \textbf{Solution for type I}

\p For sine-Gordon a solution was found some time ago by
Konik and LeClair \cite{kl1997}.
To describe their result, recall first how the Zamolodchikov
S-matrix \cite{ZZ1979} depends on the rapidity variables
{$\theta$}
and the bulk
coupling {$\beta$} via
$$ x=e^{\gamma\theta},\ q=e^{i\pi\gamma},\
\gamma=\frac{8\pi}{\beta^2}-1;$$
 it is also useful to define the variable
 $${Q=e^{4\pi^2 i/\beta^2}=
\sqrt{-q}}.$$ Then, the K-L solution, written as a two by two
matrix carrying the soliton labels, has the form
\begin{equation*}
T_{a\alpha}^{b\beta}(\theta)=f(q,x)\left(
\begin{array}{cc}
Q^\alpha\, \delta_\alpha^\beta & q^{-1/2}e^{\gamma(\theta-\eta)}
\,\delta_\alpha^{\beta-2} \\
q^{-1/2}\,e^{\gamma(\theta-\eta)}
\,\,\delta_\alpha^{\beta+2} & Q^{-\alpha}\,\delta_\alpha^\beta \\
\end{array}
\right)
\end{equation*}
where {$f(q,x)$} is not uniquely determined but, for a unitary
transmission matrix should satisfy
\begin{eqnarray*}
      \bar f(q,x)=f(q,qx),\qquad
      f(q,x)f(q,qx)=\left(1+e^{2\gamma(\theta-\eta)}\right)^{-1}.
    \end{eqnarray*}
The K-L `minimal' solution to these requirements has the
following form
$$ {f(q,x)=\frac{e^{i\pi(1+\gamma)/4}}{1+ie^{\gamma(\theta-\eta)}}\,
\frac{r(x)}{\bar r(x)}},$$
where it is convenient to put {$z=i\gamma(\theta-\eta)/2\pi$}
and
$${r(x)=\prod_{k=0}^\infty\frac{\Gamma(k\gamma+1/4 -z)
\Gamma((k+1)\gamma +3/4-z)}
{\Gamma((k+1/2)\gamma+1/4 -z)\Gamma((k+1/2)\gamma +
3/4-z)}.}$$
Remarks ({taking $\theta>0$}): it is tempting to suppose
{$\eta$} (possibly renormalized)
is essentially the same parameter as appeared in the type I
model of a defect, given the apparently similar features to
the classical soliton-defect scattering \cite{bcz2005}. In
particular:
\begin{itemize}
\item {$\eta<0$} - the off-diagonal entries dominate;
\item {$\theta>\eta>0$} - the off-diagonal entries dominate;
\item {$\theta< \eta>0$} - the diagonal entries dominate.
\end{itemize}
The particular rapidity {$\theta=\eta$}, which had a special
meaning in the classical picture, is not so special here.
Nevertheless, the function $f(q,x)$ has a simple pole nearby
at $${\theta=\eta-\frac{i\pi}{2\gamma} \rightarrow \eta,\
\beta\rightarrow 0},$$
and this pole is like a resonance, with complex energy,
$${E=m_s\cosh\theta=m_s(\cosh\eta\cos(\pi/2\gamma) -
i\sinh\eta\sin(\pi/2\gamma))}.$$
Clearly, it has a `width' proportional to {$\sin(\pi/2\gamma)$},
which tends to zero as $\beta\rightarrow 0$. The fact solitons
are treated differently to anti-solitons in the diagonal
entries of the K-L solution also has its origins in the
Lagrangian piece linear in time derivatives, a feature seen
in semi-classical arguments.

\p The Zamolodchikov S-matrix has `breather' poles
corresponding to
soliton-anti-soliton bound states at
$${\Theta =i\pi(1-n/\gamma),\ n=1,2,...,n_{\rm max}<\gamma};$$
and  the bootstrap can be used to calculate the transmission
factors for breathers. For the lightest
 breather the result is:
$${T(\theta)=-i\frac{\sinh\left(\frac{\theta-\eta}{2}-
\frac{i\pi}{4}\right)}
{\sinh\left(\frac{\theta-\eta}{2}+\frac{i\pi}{4}\right)}},$$
which has precisely the form of the transmission factor
for a plane wave encountering a defect in the linearised
sine-Gordon model.

\p \textbf{Solution for type II}

\p Apart from the solution obtained by Konik and LeClair,
there is another, more general, set of solutions to the
quadratic compatibility relations for the transmission
matrix \cite{cz2010}:
\begin{equation}\nonumber
 T_{a\alpha}^{b\beta}(\theta)=\rho(\theta)\,\left(%
\begin{array}{ccc}
 (a_+Q^{\alpha}+a_-Q^{-\alpha}\, x^2)\,
 \delta^{\beta}_{\alpha} &
  x\,(b_+Q^{\alpha}+b_-Q^{-\alpha})\,
  \delta^{\beta-2}_{\alpha} \\
  x\,(c_+ Q^{\alpha} + c_-Q^{-\alpha})\,
  \delta^{\beta+2}_{\alpha} & (d_+Q^{\alpha}\,x^2+d_-
  Q^{-\alpha} )\,\delta^{\beta}_{\alpha} \\
\end{array}%
\right)
\end{equation}
where {$x=e^{\gamma\theta}$} and
the free constants satisfy the two constraints
$${a_\pm\, d_\pm - b_\pm \,c_\pm=0}.$$
These and {$\rho(\theta)$} are constrained further by
crossing and unitarity. More details are provided in
ref\cite{cz2010} but a few remarks are in order:
\begin{itemize}
\item for a choice of parameters similarity with classical
scattering suggests this descibes a type II defect;
\item with $a_-=d_+=0$ and $b_+=c_-=0$ or $b_-=c_+=0$
(and after a similarity transformation), it reduces to
the type I solution;
\item for certain other choices of parameters it reduces
to a direct sum of the Zamolodchikov S-matrix itself and
two infinite dimensional pieces, suggesting that a type
II defect is itself similar to a soliton.
\end{itemize}

\textbf{Solution for the Tzitz\'eica model}

\p Recall:  the Tzitz\'eica quantum field theory is a
little different to its classical version in so far as
it possesses mass-degenerate solitons with charges
$\pm 1$ and $0$ whose scattering is described by the
Izergin-Korepin-Smirnov S-matrix \cite{iks1990}.
Using the latter and solving the compatibility relations
leads to \cite{cz2011}
\begin{equation*}
T_{a\alpha}^{b\beta}(\theta)= \rho(\theta)\left(
  \begin{array}{ccc}
    (\varepsilon^2\,q^{2\,\alpha}+\tau^2 \,q^{-2\,\alpha}\,x)\,
    \delta^{\beta}_{\alpha}& \varepsilon\mu(\alpha) \,
    \delta^{\beta-1}_{\alpha} & M(\alpha)\,
    \delta^{\beta-2}_{\alpha} \\
    \tau\lambda(\alpha)\,x\,\delta^{\beta+1}_{\alpha}&
    (\tilde{\tau}\,\varepsilon+\tau\,\tilde{\varepsilon}\,x)\,
    \delta^{\beta}_{\alpha} & \tilde{\tau}\,\,
    \mu(\alpha) q^{-2\,\alpha-1} \,\delta^{\beta-1}_{\alpha} \\
    L(\alpha)\,x\,\delta^{\beta+2}_{\alpha}&
    \tilde{\varepsilon}\,q^{2\,\alpha-1}\,\lambda(\alpha)
    \,x \,\delta^{\beta+1}_{\alpha} &
    (\tilde{\varepsilon}^2\,q^{2\,\alpha}\,x+
    \tilde{\tau}^2 \,q^{-2\,\alpha})\,
    \delta^{\beta}_{\alpha} \\
  \end{array}
\right)
\end{equation*}
where
\begin{eqnarray*}M(\alpha)=\mu(\alpha)\,\mu(\alpha+1)\,
    \frac{q^{-2\,\alpha-1}}{1+q^{2}}, \quad L(\alpha)=
    \lambda(\alpha)\,\lambda(\alpha-1)\,\frac{q^{2\,
    \alpha-1}}{1+q^2},\ \ \\
\phantom{M}\mu(\alpha)\,\lambda(\alpha+1)=(q+q^{-1})\,
\left(\tau\,\tilde{\tau}\,q^{-2\,\alpha-1}
+\varepsilon\,\tilde{\varepsilon} \,q^{2\,\alpha+1}
\right),\ q=e^{i 4\pi^2/\beta^2}.
\end{eqnarray*}
As before, the overall factor and some combinations
of constants are constrained  by insisting on crossing
and a consistent bootstrap (in this case the model
does not possess unitary scattering but each soliton is
a bound state of two others). Thus,
\begin{eqnarray*} \rho(\theta)\,\rho(\theta
i\pi)\ (\tilde{\tau}\varepsilon+\tau\tilde{\varepsilon}q^{-4}x)
\,(\tilde{\tau}\varepsilon-\tau\tilde{\varepsilon}
q^{-2}x)=1,\ \ %\\\nonumber
\rho(\theta)=(\tilde{\tau}\varepsilon+\tau\tilde{\varepsilon}x)\,
\rho(\theta+i\pi/3)\,\rho(\theta-i\pi/3).
\end{eqnarray*}
This solution looks complicated but it can be approached
in an alternative manner suggested by Weston \cite{w2010}
and used in the sine-Gordon case to reproduce the type
II transmission matrix described above. The idea is to
associate the transmission matrix with an intertwiner
linking the two co-products of a finite dimensional
and an infinite-dimensional representation of a Borel
subalgebra of the `quantised' affine algebra
$U_q(a_2^{(2)})$. The solitons belong to the finite
dimensional representation while the infinite-dimensional
representation describes the defect. Moreover it is
convenient to describe the infinite dimensional
representation in terms of operators that create and
annihilate topological charge. First, a short
introduction to the algebra might be useful.

\p With $\alpha_0$ the shorter root, the Cartan
matrix of $a_2^{(2)}$ is
\begin{equation*}\label{CartanMa22}
C_{ij}=\left(
  \begin{array}{cc}
    \phantom{*}2 & -1 \\
    -4 & \phantom{*}2 \\
  \end{array}
\right),\quad i,j=0,1.
\end{equation*}
 The quantised algebra $U_q(a_2^{(2)})$ introduced
 by Drinfel'd and Jimbo (see \cite{j1994}) has
 six generators $$\{X^{\pm}_1,
 X^{\pm}_0, \ K_1, K_0\},$$ satisfying:
\begin{eqnarray*}\label{uqa22comrel}
[K_1,K_0]=0,\quad K_i\,K_i^{-1}=K_i^{-1}\,K_i=1,\quad
i=0,1,\phantom{mmml}\nonumber\\
K_0\,X_0^{\pm}\,K_0^{-1}=q^{\pm 4}\,X_0^{\pm},\quad
[X_0^+,X_0^-]=\frac{K_0^2-K_0^{-2}}{q^4-q^{-4}},
\phantom{mmmll}\nonumber\\
K_1\,X_1^{\pm}\,K_1^{-1}=q^{\pm 1}\,X_1^{\pm},
\quad
 [X_1^+,X_1^-]=\frac{K_1^2-K_1^{-2}}{q-q^{-1}},
 \phantom{mmml}\nonumber\\
K_1X_0^{\pm}\,K_1^{-1}=q^{\mp 2}\,X_0^{\pm},
\quad K_0\,X_1^{\pm}\,K_0^{-1}=q^{\mp 2}\,X_1^{\pm},
\quad [X_1^{\pm},X_0^{\mp}]=0,
\end{eqnarray*}
and the Serre relations,
\begin{eqnarray*}\label{uqa22Serrerelations}
\nonumber&&\sum_{k=0}^5 (-1)^k
\begin{bmatrix}
5 \\
k
\end{bmatrix}
_q
\left(X_1^{\pm}\right)^{5-k} X_0^{\pm} \left(X_1^{\pm}
\right)^{k}=0,\\
&&\sum_{k=0}^2 (-1)^k
\begin{bmatrix}
2 \\
k
\end{bmatrix}
_{q^4}
\left(X_0^{\pm}\right)^{2-k} X_1^{\pm} \left(X_0^{\pm}
\right)^{k}=0.
\end{eqnarray*}
%\end{frame}
%\begin{frame}
The coproducts $\Delta,\ \Delta^\prime$ are given by:
\begin{equation*}
\Delta(K_i)=K_i\otimes K_i,\quad \Delta(X^{\pm}_i)
=X_i^{\pm}\otimes K_i^{-1}+K_i\otimes X_i^{\pm},
\quad i=0,1,
\end{equation*}
and
$$\Delta'(K_i)=\Delta(K_i),\quad \Delta'(X^{\pm}_i)
=K_i^{-1}\otimes X_i^{\pm} +X_i^{\pm} \otimes K_i,
\quad i=0,1.$$
The fundamental representation of this algebra is
three-dimensional
and the intertwiner between the two co-products of
three-dimensional representations is the S-matrix.

\p For our purposes, an infinite-dimensional representation
of the Borel subalgebra generated by $\{X_1^+,\,X_0^+,\,K_1,\,K_0\}$
is also required and conveniently realised using a pair of
annihilation and creation operators similar to those introduced
by  Macfarlane and Biedenharn twenty years ago \cite{mb1989}.
In detail, they are defined by
$$a|j\rangle =F(j)\,|j-1\rangle ,\quad \hat{a}|j\rangle =
|j+1\rangle ,\quad N|j\rangle =j|j\rangle ,\quad j
\in \mathbb{Z}.$$
Note, $F(j)$ need not vanish for any $j$, since $j$
represents a topological charge. Also,
$$a\hat{a}=F(N+1),\ \hat{a} a=F(N),\ aG(N)=G(N+1)a,\
\hat{a} G(N)=G(N-1)\hat{a}$$
where $G(N)$ is any function of the number operator.
It is not necessary to insist on the conjugation
relation $\hat{a}=a^\dagger$.
Using these, the generators of the Borel subalgebra
are taken to be:
$$X_1^+=\hat{a},\quad X_0^+=a \,a,\quad K_1=
\kappa_1\, q^{N},\quad K_0=\kappa_0\,q^{-2N},$$
where $\kappa_0$ and $\kappa_1$ are constants.
This choice satisfies the Borel sub-algebra and
the Serre relations require,
\begin{eqnarray*}
\nonumber\sum_{k=0}^5 (-1)^k
\begin{bmatrix}
5 \\
k
\end{bmatrix}
_q
 F(N+k)F(N+k+1)=0,\\
\sum_{k=0}^2 (-1)^k
\begin{bmatrix}
2 \\
k
\end{bmatrix}
_{q^4}
 F(N+2k)=0,\phantom{mm}
\end{eqnarray*}
%\end{frame}

%\begin{frame}
which in turn require
\begin{eqnarray*}
\hat{a}\,a=F(N)=\left(b_1\,(-)^N+c_1\right)
\,q^{-2 N}+\left(b_2\,(-)^N+c_2\right)\,q^{2 N}\\
b_1\,c_2=b_2\,c_1.\phantom{mmmmmmmmmmmm}
\end{eqnarray*}

%\pause
Finally, it is necessary to use a homogeneous
gradation \cite{e1993}, namely
\begin{equation*}
E_1=X^{+}_1,\quad F_1=X^{-}_1,\quad E_0=x\,X^{+}_0,
\quad F_0=x^{-1}\,X^{-}_0.
\end{equation*}
 Then the transmission matrix $T$ is an intertwiner
 of the infinite dimensional representation with space
 $\mathcal{V}$ and the three-dimensional
 representation with space $V$:
$$T(z/x):\mathcal{V}_{z}\otimes V_{x}\rightarrow
\mathcal{V}_{z}\otimes V_{x}$$
achieved by solving the linear condition (for any
element $b$ of the Borel subalgebra),
\begin{equation*}\label{lineareqforT}
T\,\Delta(b)=\Delta'(b)\,T,
\end{equation*}
to find
%\end{frame}
%\begin{frame}
\begin{equation*}\label{lineareqTa22}
T=\left(
  \begin{array}{ccc}
    a'\,q^{-2N}+a''\,q^{2N} & k \,q^{N}\,a
    & v\,a\, a \\
    j \,q^{-N}\,\hat{a} & b & i \,q^{-N}\,a \\
    w\, \hat{a} \,\hat{a} & l \,q^{N}\,\hat{a} &
    c'\,q^{2N}+c''\,q^{-2N} \\
  \end{array}
\right)
\end{equation*}
where the coefficients have to be chosen suitably
\cite{cz2011}.

\vspace{10pt}
\textbf{\large 5\quad Discussion}

\p There is no reason why defects of type I or type II
should be static. In fact, for type I, travelling defects
have been analysed within the sine-Gordon theory
\cite{bcz2005}, and, if there are several moving at
different speeds their classical scattering is
consistent with their interactions with solitons
owing to the `permutability theorem' for repeated
B\"acklund transformations. On the other hand, their
quantum scattering is not yet completely determined,
though there is a
candidate S-matrix compatible with the soliton
transmission matrix \cite{bcz2005, w2010, bgknr2010}.
There are many interesting features pertaining to
sine-Gordon defects that have been examined by other
people, for example refs\cite{ hk2008, bs2008}, and
also to defects that might occur within extensions
to sine-Gordon including fermions or analysing the
relationship with the massive Thirring model \cite{gyz2006},
or within other integrable field theories not of Toda
type, such as NLS, KdV or mKdV \cite{cz2006}.
Defects can be constructed within the {$a_r^{(1)}$}
series of affine Toda models \cite{bcz2004}  and
transmission matrices have been written down for
both types I and II \cite{cz2007,cz2011}.
Generalised annihilation and creation operator
representations of Borel subalgebras for all affine
quantum groups can be constructed \cite{cz2011},
though there are alternatives arising in other
applications: for example, to construct representations
of Baxter's $Q$-operator in the context of solvable
spin chains \cite{bhk2002}.

\vspace{10pt}

\textbf {\large Acknowledgements}

\p I would like to thank the conference organisers for
the opportunity to present this work, Peter Bowcock
and Cristina Zambon for a long and enjoyable collaboration,
Robert Weston for several illuminating conversations,
and  the UK Engineering and Physical Sciences
Research Council for its
support via the grant EP/F026498/1.

\end{document}